\documentclass[12pt]{article}
\usepackage{amsmath}
\usepackage{amsfonts}
\usepackage{amssymb}

\numberwithin{equation}{section}

\newcommand\sbr[2]{\left\lbrack\,{#1}\, ,\,{#2}\,\right\rbrack} 
\def\rf#1{(\ref{eq:#1})}
\def\lab#1{\label{eq:#1}}
\def\bj{{\bar J}}
\def\sj{{\jmath}}
\def\bsj{{\bar \jmath}}
\def\p{\phi}
\providecommand*{\pder}[3][]{%
\frac{\partial^{#1}#2}{\partial #3^{#1}}}
\providecommand*{\dder}[3][]{%
\frac{d^{#1}#2}{d #3^{#1}}}
\providecommand*{\iu}%
{\ensuremath{\mathrm{i}\,}}

\begin{document}
\vspace*{-1cm}
\noindent
\vskip .3in

\begin{center}
{\large\bf Darboux-B\"acklund Derivation of
Rational Solutions\\ of the Painlev\'e IV Equation }
\end{center}
\normalsize
\vskip .4in

\begin{center}
H. Aratyn

\par \vskip .1in \noindent
Department of Physics \\
University of Illinois at Chicago\\
845 W. Taylor St.\\
Chicago, Illinois 60607-7059\\
\par \vskip .3in

\end{center}

\begin{center}
J.F. Gomes and A.H. Zimerman

\par \vskip .1in \noindent
Instituto de F\'{\i}sica Te\'{o}rica-UNESP\\
Rua Pamplona 145\\
01405-900 S\~{a}o Paulo, Brazil
\par \vskip .3in

\end{center}

\begin{center}
{\large {\bf ABSTRACT}}\\
\end{center}
\par \vskip .3in \noindent

Rational solutions of the Painlev\'e IV equation are constructed
in the setting  of  pseudo-differential Lax formalism  describing
AKNS hierarchy subject to the additional non-isospectral Virasoro 
symmetry constraint.
Convenient Wronskian representations for rational solutions 
are obtained by  successive actions of the Darboux-B\"acklund 
transformations.
%
\section{Introduction}
Two main observations are put together in this paper with the purpose
of developing a systematic method rooted in
Darboux-B\"acklund techniques within the pseudo-differential Lax formalism
to generate  rational solutions of the Painlev\'e IV equation.
First, we observe that the AKNS hierarchy subject to the additional
non-isospectral Virasoro symmetry constraint (also known as the string equation)
reduces to the Painlev\'e IV equation.
Secondly, we make use of the fact that
the  Darboux-B\"acklund (DB) transformations commute 
with the additional-symmetry Virasoro 
flows  \cite{Aratyn:1996nb} and thus these transformations
can effectively be used to construct all known rational solutions 
\cite{Clarkson-jmp,Clarkson,kajiwara,noumi} of 
the Painlev\'e IV equation \rf{painy} out of a few basic ones.

\section{AKNS hierarchy with string condition}
To introduce the AKNS Lax formalism with additional 
Virasoro symmetry flows we follow \cite{Aratyn:1996nb}
and define the pseudo-differential Lax hierarchy 
as described by a dressing formula
\begin{equation}
L = W \partial_x W^{-1}= \partial_x-r\partial_x^{-1}q,
\quad W= 1 + \sum_1^{\infty} w_n \partial_x^{-n}
\lab{f-5}
\end{equation}
The associated $t_n$-flows of this hierarchy:
\begin{equation}
\pder{r}{t_n} = L^{n }_{+} (r ) \qquad; \qquad
\pder{q}{t_n} = - \left( L^{\ast} \right)^{n }_{+} (q )
\quad\;\; n=1,2, \ldots
\lab{eigenlax}
\end{equation}
reproduce for $n=2$ the AKNS equations :
\begin{equation}
\pder{}{t_2} q +q_{xx}-2 q^2 r = 0, \qquad
\pder{}{t_2} r -r_{xx}+2 q r^2 = 0 \, .
\lab{akns}
\end{equation}
Furthermore, it is useful to introduce as in  \cite{squared}
a squared eigenfunction $\rho$  such that 
\[
\rho_x   = - 2 r q
\]
and, which in particular, satisfies \cite{squared} :
\begin{equation}
\pder{}{t_2} \rho = 2 \left( -q r_x+r q_x\right)\, .
\lab{aknsd}
\end{equation}
The above formalism is augmented by the Orlov-Schulman 
operator \cite{add-symm} :
\[
M =  W \left(  \sum_{l \geq 1} l t_l \partial_x^{l-1} \right) W^{-1}
\]
conjugated to the Lax operator according to 
the commutation relation 
$\left\lbrack L \, , \, {M} \right\rbrack=1$.

The additional Virasoro symmetry flows are then defined as
\begin{equation}
{\bar \partial}_{k,n} L = - \left\lbrack{\left( M^n L^k\right)_{-}}
\, , \, {L}\right\rbrack =
\left\lbrack {\left( M^n L^k\right)_{+}}  \, , \, {L}\right\rbrack  
+ n M^{n-1} L^k  
\lab{add-symm-L}
\end{equation}
and they commute with the usual isospectral flows as defined
by $\partial L /\partial {t_n} = [L^{n }_{+}\,,\, L]$ or
\rf{eigenlax}.

For $n=1$ formula \rf{add-symm-L} yields
\begin{equation}
\left( {\bar \partial}_{k,1} L\right)_{-} =
{\sbr{\left( M L^k \right)_{+}}{L}}_{-} + \left( L^k  \right)_{-}\, ,
\lab{symm1}
\end{equation}
which can be rewritten as
\begin{equation}
\left( {\bar \partial}_{k,1} L\right)_{-} =  \left( M L^k \right)_{+} (r) \partial^{-1}
q -  r \partial^{-1} \left( M L^k \right)^{\ast}_{+} ( q )  +
\sum_{j=0}^{k-1} L^{k-j-1} (r) \partial^{-1}
\left( L^{\ast}\right)^{j} ( q ) \, .
\lab{pak1}
\end{equation}
For $k=0,1,2$ these  flows form the $sl(2) $ subalgebra of the 
Virasoro algebra and preserve the form of  the Lax operator of the 
AKNS hierarchy. Subsequently, they can  equivalently be 
described by their action on the eigenfunction $r$ and
the adjoint eigenfunction $q$ :
\begin{equation}
\left(\partial_{\tau} L\right)_{-} = \left(\partial_{\tau} r\right) \partial^{-1} q + r
\partial^{-1} \left( \partial_{\tau} q\right)
\lab{addflo}
\end{equation}
with $\partial_\tau \equiv {\bar \partial}_{k,1}\; (k=0,1,2)$.
Explicitly, one finds \cite{Aratyn:1996nb} :
\begin{xalignat}{2}
{\bar \partial}_{0,1} r &= \left( M\right)_{+}  (r), &
{\bar \partial}_{0,1} q &= - \left( M\right)^{\ast}_{+} ( q  ), \lab{papsiz} \\
{\bar \partial}_{1,1} r &= \left( ML\right)_{+} (r) + (1-\nu) r, &
{\bar \partial}_{1,1} q &= - \left( ML\right)^{\ast}_{+} (q) + \nu q,
\lab{papsi1} \\
{\bar \partial}_{2,1} r &= \left( M L^2 \right)_{+} (r) + L (r),&
{\bar \partial}_{2,1} q &= - \left( M L^2 \right)^{\ast}_{+} (q) + L^{\ast}
(q).
\lab{papsi2}
\end{xalignat}
Note arbitrariness in equation \rf{papsi1} 
expressed by a parameter $\nu$.

In what follows we impose the symmetry conditions:
\begin{equation}
{\bar \partial}_{1,1} r = 0 , \quad
{\bar \partial}_{1,1} q= 0 \, .
\lab{cond11}
\end{equation}
Since 
\begin{gather*}
ML= W \left(  \sum_{l \geq 1} l t_l \partial_x^{l} \right) W^{-1},\quad
\left( ML\right)^{\ast} =  W^{-1\, \ast} \left( - \partial_x x +\partial_x^2 2 t_2 +{\ldots} \right)
 W^{ \ast}\\= -1 +W^{-1\, \ast} \left( - x\partial_x +2 t_2 \partial_x^2  +{\ldots} \right)
 W^{ \ast}
\end{gather*}
it follows that
\[
\left( ML\right)_{+} = x \partial_x + 2 t_2  \left( L \right)^{2 }_{+}+ {\ldots} ,\quad
-\left( ML\right)^{\ast}_{+}= 1 +x \partial_x - 2 t_2   \left( L^{\ast} \right)^{2 }_{+}+ {\ldots} 
\]
Thus in view of equation \rf{papsi1} the condition \rf{cond11} 
amounts to 
\begin{equation}
\begin{split}
- x q_x -2 t_2 \pder{}{t_2} q &= q + \nu q, \, \;\;\;
 x r_x +2 t_2 \pder{}{t_2} r = -r + \nu r \\
 x \rho_x +2  t_2 \pder{}{t_2} \rho &= - \rho \, .
\end{split}
\lab{string_eq3}
\end{equation}
The model we consider is obtained by applying
the string equation \rf{string_eq3}
to eliminate $t_2$-dependence from $r, q, \rho$
and setting the higher flows to zero.
Setting $t_2=-1/4$ and eliminating $t_2$-flow dependence 
from eq. \rf{string_eq3} by 
inserting values from \rf{akns}-\rf{aknsd} yields :
\begin{equation}
\begin{split}
- x q_x +\frac12 \left(-q_{xx}+2q^2r\right) &= q + \nu q ,\quad
 x r_x - \frac12 \left(r_{xx}-2q r^2\right) = -r + \nu r \\
 \rho+ x \rho_x &= \rho-2 x rq = q_x r-q r_x
\end{split}
\lab{string_eq4}
\end{equation}
Multiplying the first equation by $r_x$ and second by $q_x$
and adding them one obtains:
\[
\left(q_x r_x - q^2 r^2 \right)_x = 2 (q_x r-q r_x) -2 \nu (qr)_x
\]
or after integration :
\begin{equation}
q_x r_x = q^2 r^2 +2 x \rho -2 \nu rq -(\mu^2-\nu^2)\, ,
\lab{fixc}
\end{equation}
where we have set the integration constant to $\mu^2-\nu^2$, 
with $\mu$ being an extra parameter. In what follows $\mu$
will, in addition to $\nu$, parametrize solutions of 
the Painlev\'e IV equation.
The origin of $\mu$ is very transparent in an algebraic approach
to integrable models subject to the scaling condition \cite{PVI}.

Dividing the first of eqs. \rf{string_eq4} by $q$ and second
by $r$ and summing them yields:
\[
\begin{split}
2 \nu&= x \left(-\frac{q_x}{q}+\frac{r_x}{r} \right)-\frac12 \left(
\frac{(rq)_{xx}}{rq}
- 2 \frac{r_xq_x}{rq}\right)+ 2 qr =\\
&=x \left(2x -\frac{\rho}{rq} \right)-\frac12 \left( \frac{(rq)_{xx}}{rq}
- 2 \frac{r_xq_x}{rq}\right)+ 2 qr \, , 
\end{split}
\]
where in the last line we used the third of eqs. \rf{string_eq4}.
Inserting $q_x r_x$ from eq. \rf{fixc} produces after 
multiplication by $rq$ :
\[
2 x^2 rq +x \rho - \frac12 (rq)_{xx} - 4 \nu rq +3 (rq)^2=
\mu^2-\nu^2 \, .
\]
Inserting in the above $rq=-\rho_x/2$ results in an
equation entirely expressed in terms of only one variable $\rho$ 
:
\begin{equation}
-x^2 \rho_x +x \rho +\frac14 \rho_{xxx} 
+2 \nu \rho_x + \frac34 \rho^2_x =\mu^2-\nu^2 \, .
\lab{rho-eq}
\end{equation}
One recognizes in the above equation the special case
of the Chazy I equation \cite{cosgrove}.
After multiplication of eq. \rf{rho-eq} by $\rho_{xx}$
we can rewrite the resulting equation as a total derivative,
which after an integration becomes
\begin{equation}
 \rho_{xx}^2 = 4 \left(x \rho_x-\rho\right)^2 -2 \rho_x^3
 -8 \nu \rho_x^2 +8 (\mu^2-\nu^2) \rho_x-8 C\, ,
\lab{jmo}
\end{equation}
with $C$ being an unknown integration constant. Another well-known
form  of this equation is \cite{jimbo} :
\[
 \rho_{xx}^2 = 4 \left(x \rho_x-\rho\right)^2 -2 \prod_{i=1}^3
 \left(\rho_x +v_i\right) \, .
\]
By comparison with eq. \rf{jmo} one obtains
\[
v_1+v_2+v_3=4 \nu, \quad v_1v_2+v_1v_3+v_2v_3= -4 (\mu^2-\nu^2)
, \quad v_1v_2v_3= 4 C \, .
\]
If we set the integration constant $C$ to zero then eq. \rf{jmo}
can be simplified as follows : 
\begin{equation}
 \rho_{xx}^2 = 4 \left(x \rho_x-\rho\right)^2 -2 \rho_x \big\lbrack\rho_x-2
 (\mu-\nu)\big\rbrack
\big\lbrack\rho_x+2 (\mu+\nu)\big\rbrack
\lab{jmo1}
\end{equation}
or in a slightly different and more  convenient form :
\begin{equation}
\bigl( 2 \left(x \rho_x-\rho\right)+\rho_{xx}\bigr)\bigl( 
2 \left(x \rho_x-\rho\right)-
\rho_{xx}\bigr) = 2 \rho_x  \big\lbrack\rho_x-2
 (\mu-\nu)\big\rbrack
\big\lbrack\rho_x+2 (\mu+\nu)\big\rbrack \, .
\lab{jmo2}
\end{equation}
First, one notes that equation \rf{jmo2} is manifestly invariant 
under $\mu \to -\mu$ and thus a solution of equation \rf{jmo2}
with one value of $\mu$ solves this equation for $-\mu$ as well.
Next, one notes that the left hand side of equation \rf{jmo2}
remains invariant under substitution $\rho={\widetilde \rho} + 
k x$ with $k$ being a constant.
For $k=2 (\mu-\nu)$ and $k=-2 (\mu+\nu)$ the right hand side
can be given the form $2 {\widetilde \rho}_x  \lbrack {\widetilde \rho}_x-2
({\widetilde \mu}-{\widetilde \nu})\rbrack \lbrack{\widetilde \rho}_x+
2 ({\widetilde \mu}+ {\widetilde \nu})\rbrack$ with
${\widetilde \nu} = 3 \mu/2-\nu/2$, ${\widetilde \mu}= \pm(\mu+\nu)/2$
and ${\widetilde \nu} = -3 \mu/2-\nu/2$, ${\widetilde \mu}= \pm(\mu-\nu)/2$,
respectively. Hence for these values of $k$ the transformation $\rho \to {\widetilde \rho}$
takes the ``old'' solution of equation \rf{jmo2} to the ``new'' solution of 
equation \rf{jmo2} with the new parameters  ${\widetilde \mu}$,
${\widetilde \nu}$.
\section{Basic Polynomial Solutions}
One finds by inspection that there exists a class of polynomial solutions 
of equation \rf{jmo2} for $\rho$ being of the 
form $\rho=bx^3 +c x$.
With both coefficients, $b$ and $c$, being non-zero
the polynomial solution is :
\begin{equation}
\rho = \frac{8}{27} x^3 \pm \frac{4}{3} x, \quad \mu^2 = \frac19,
\;\; \nu = \mp 1\,.
\lab{rho13}
\end{equation}
For $c=0$, one finds that the unique non-zero solution
requires $b=8/27$  with  $\nu=0, \mu^2 =4/9$ :
\begin{equation}
\rho = \frac{8}{27} x^3 , \quad \mu^2 = \frac49,
\;\; \nu = 0\, .
\lab{rhob1}
\end{equation}
Finally, one notes that setting $b=0$ causes both 
$x \rho_x-\rho$ and $\rho_{xx}$ to vanish together
with the left hand side of eq. \rf{jmo2}. 
Thus $\rho=c x$ is a solution to eq. \rf{jmo2}
for three values of $c$, namely $c= 0, 2  (\mu-\nu) , 
- 2 (\mu+\nu)$ for which the right hand side of eq. \rf{jmo2}
vanishes as well.

In a text below, the polynomial solutions of equation \rf{jmo2} 
will serve as seeds of chains of the Darboux-B\"acklund transformations.

For $\rho=c x$ the product $rq$ is a constant, $-c/2$.
In addition, $q_x r-q r_x  = \rho-2 x rq =2 c x$ and thus
$r q_x= c x $, $r_x q=- c x $.
Multiplying eq. \rf{fixc} by $r$ and then by $q$ one gets
{}from $r q_x= c x $, $r_x q=- c x $ that
\[
c x r_x = \left( 2 c x^2 +c^2/4+c \nu -(\mu^2-\nu^2)\right)r, \;\;\,
-c x q_x =\left( 2 c x^2 +c^2/4+c \nu -(\mu^2-\nu^2)\right)q
\]
For $c= 2  (\mu-\nu) , - 2 (\mu+\nu)$ the above equations
simplify to 
\[x r_x = 2 x^2 r, \;\; - x q_x=2 x^2 q\;\;,\;
\to\;\; r=k_{1} e^{x^2},\;\;q=k_{- 1} e^{-x^2}
\]
and as long $c=-2 k_{1} k_{- 1} \ne0$ the above solutions do not lead to 
non-zero solutions of the Painlev\'e IV equation.

We now consider the case of $b=c=0$. Then $rq=0$ with either $r=0$ or 
$q=0$ and it follows  from \rf{fixc} that $\mu^2=\nu^2$.

Another useful expression for eqs. \rf{string_eq4} is :
\begin{equation}
\left(e^{-x^2}r_x\right)_x +2 (\nu-1)e^{-x^2}r   = 2 (rq)e^{-x^2} r, \quad
 \left(e^{x^2}q_x\right)_x + 2 \left(\nu+1\right)e^{x^2} q = 2 (rq)e^{x^2} q
\lab{akns3}
\end{equation}
For $q=0$ and $\nu -1 =n $ being a positive (or zero) integer the first
equation of eqs. \rf{akns3} is 
Hermite's equation, alternatively known as $r_{xx}-2 x r_x +2 
n r=0$ with $r(x)=H_{\nu-1} (x) $ being a solution.
Thus one obtains, $r=H_0=1$ for $\nu=1$, $r=H_1=2x$ for $\nu=2$
(the solution we have already studied) and $r=H_2=4x^2-2$ for $\nu=3$ etc.

For $-\nu \ge 0$ being a positive integer a
substitution
\[
r(x) = e^{x^2} \, g(x)
\]
leads to the following equation
\[
g_{xx}+2 x g_x +2 \nu g= g_{xx}+2 x g_x -2 m g=0, \;\;
m=-\nu =0,1,2,{\ldots} 
\]
which has a polynomial solution 
\begin{equation} 
g_m (x)={\widehat H}_m (x) = e^{-x^2} \dder[m]{}{x} e^{x^2}= (- \iu)^m H_m (\iu x)
\lab{hathermit}
\end{equation}
with  $g_{m=0}=1$,  $g_{m=1}=2x$, $g_{m=2}=2(1+2x^2)$
and $g_{m=3}=(3+2x^2) 4 x$ etc. 
To summarize :
\begin{equation}
r(x) =  
\begin{cases} 
  H_{\nu-1},  & \mbox{if }\nu >0 \mbox{ is integer} \\
 e^{x^2} {\widehat H}_{-\nu} ( x), & \mbox{if }\nu \le 0 \mbox{ is integer} 
\end{cases}
\lab{rhermit}
\end{equation}
for $q=0$.
Substituting 
\[
q= e^{-x^2} \, f(x) , \qquad 
\]
into \rf{akns3} with $r=0$ one gets the following equation
for $f$ :
\[
f_{xx}-2 x f_x +2 \nu f=0, 
\]
which is a Hermite's equation for $H_n (x)$
for $\nu=n\ge 0$. Thus $f(x) = H_{\nu} (x)=H_n (x)$ with
$n=0,1,2,{\ldots} $ and 
$q (x) = \exp (-x^2) H_{\nu} (x)$ as 
long as $\nu$ is a positive integer. 
For $\nu+1=-m$, where $m=0,1,2,{\ldots} $ the equation for $q$,
\[
q_{xx}+2xq_x-2mq=0
\]
is identical to equation for $g$ written above and 
has the same solutions.
To summarize:
\begin{equation}
q(x) =  
\begin{cases} 
e^{-x^2}  H_{\nu},  & \mbox{if }\nu \ge 0 \mbox{ is integer} \\
 {\widehat H}_{-\nu-1} (x), & \mbox{if }\nu < 0 \mbox{ is integer} 
\end{cases}
\lab{qhermit}
\end{equation}
for $r=0$.

\section{Painlev\'e IV equation and its Hamiltonian formalism}
In terms of variables:
\begin{equation}
y = - \frac{q_x}{q}- 2 x , \quad w = \frac{r_x}{r}- 2 x,
\quad z = -2 rq +2 \left(\mu +\nu\right)
\lab{ywz}
\end{equation}
we can express equations \rf{string_eq4} as
\begin{equation}
\begin{split}
y_x &= z+y^2 +2 x y -2 \mu, \quad \;
w_x = -z -w^2-2 x w+2 \mu\\
z_x&= 2 rq (y-w)= -\frac{z^2}{2w} +w z-2 (\mu+\nu) w+2 \mu \frac{z}{w}\\
&=\frac{z^2}{2y} -y z+2 (\mu+\nu) y-2 \mu \frac{z}{y}
\end{split}
\lab{pain1}
\end{equation}
Applying one more derivative on the first equation and eliminating 
$z$ yields the Painlev\'e IV equation:
\begin{equation}
y_{xx} = \frac{1}{2y} y_x^2+\frac32 y^3 +4 x y^2 +2(x^2+\nu+1)y
-2  \frac{\mu^2}{y}
\lab{painy}
\end{equation}
Repeating the same procedure for the equation with $w$  yields 
almost identical equation (note however a shift in $\nu$ as compared to
eq. \rf{painy}):
\begin{equation}
w_{xx} = \frac{1}{2w} w_x^2+\frac32 w^3 +4 x w^2 +2(x^2+\nu-1)w
-2  \frac{\mu^2}{w}
\lab{painw}
\end{equation}
As found by Okamoto \cite{okamoto},  equations \rf{pain1} for a pair of
variables $(y,z)$ possess a Hamiltonian representation.
Define namely
\begin{equation}
H = -2 P^2 Q +\left( Q^2 +2 x Q -2 \mu\right) P +\frac12 (\mu+\nu) Q \, .
\lab{hamo}
\end{equation}
Then the Hamilton equations are:
\begin{align}
Q_x &= \pder{H}{P} = -4 Q P +Q^2+2 x Q -2 \mu \lab{qx}\\
P_x &=- \pder{H}{Q} = 2 P^2 -2 QP -2 x P -\frac12 \left( \mu +\nu\right) \, .
\lab{px}
\end{align}
They agree with eqs. \rf{pain1} for 
$Q=y$ and $z=-4 Q P$ or $P =-z/4y$.
Due to the Hamilton equations it holds that
\begin{equation}
\dder{}{x} H = H_x =2 Q P= -\frac12 z = rq -\left(\nu + \mu\right)
=-\frac12 \rho_x -\left(\nu + \mu\right) \, .
\lab{hx}
\end{equation}
Thus, up to a constant,
\begin{equation}
H= - \frac12 \rho -x \left(\nu + \mu\right) \, .
\lab{hamrho}
\end{equation}
Plugging the product 
$Q P = - \left( \rho_x+2 (\mu+\nu)\right)/4$
into expression \rf{hamo} for $H$
and using eq. \rf{hamrho} one gets
\[
P \left( -rq-\mu+\nu\right) + Q \frac12 rq+ x rq =-\frac12 \rho
\]
or
\[
\frac12 P  \left(\rho_x-2 (\mu-\nu)\right)-  \frac14 Q \rho_x=
\frac12 \left(x \rho_x-\rho\right) \, .
\]
Define:
\begin{equation}
P = P_0 \frac{2}{\rho_x-2 (\mu-\nu)},\qquad
Q =-4 Q_0 \frac{1}{\rho_x} \, .
\lab{PQdefs}
\end{equation}
Then
\[
\begin{split}
P_0+Q_0 &= \frac12 \left(x \rho_x-\rho\right)\\
P_0Q_0 &=\frac{1}{32} \rho_x \left(\rho_x-2 (\mu-\nu)\right)
\left(\rho_x+2 (\mu+\nu)\right)= \frac{1}{64}\left( 4 \left(x \rho_x-\rho\right)^2-
\rho_{xx}^2\right)\\
&=\frac{1}{64}\left(2 \left(x \rho_x-\rho\right)+\rho_{xx}\right)
\left(2 \left(x \rho_x-\rho\right)-\rho_{xx}\right)
\end{split}
\]
where use was made of eq. \rf{jmo1}.
Solutions to the above equation are easily found to be 
\[
P_0 = \frac{1}{8} \left(2 \left(x \rho_x-\rho\right)\pm \rho_{xx}\right)
, \quad 
Q_0 = \frac{1}{8} \left(2 \left(x \rho_x-\rho\right)\mp\rho_{xx}\right)\, ,
\]
which gives two answers for expression for the Painlev\'e
IV solution $y=Q$ in terms of the solution to the $\rho$-equation
\rf{jmo2}
\begin{equation}
y_{\pm}=Q  =  \frac{-1}{2\rho_x}\left(2 \left(x \rho_x-\rho\right)\pm\rho_{xx}\right)
\lab{yrho}
\end{equation}
The above ambiguity in signs can be explained by comparing with
the third equation in \rf{pain1}. Consider namely the difference
$y_{+}-y_{-}$ of two solutions from eq. \rf{yrho}:
\[
y_{+}-y_{-}= - \frac{\rho_{xx}}{\rho_{x}}= \frac{z_x}{2 rq}
= y-w
\]
where we used the third of equations in \rf{pain1}.
Thus
$y_{+}= y , \; y_{-} =w$.
It is natural to connect these associations by a B\"acklund 
transformation by defining a ``barred'' system such that 
\[
w=y_{-}={\bar y}_{+}={\bar y}= \frac{-1}{2{\bar \rho}_x}\left(2 \left(x {\bar
\rho}_x-{\bar \rho}\right)+{\bar \rho}_{xx}\right)
\]
and thus the new ${\bar \rho}$ variable needs to satisfy
\[
 \frac{\rho}{\rho_x}+\frac{\rho_{xx}}{2\rho_x}=
 \frac{{\bar \rho}}{{\bar \rho}_x}-\frac{{\bar \rho}_{xx}}{2{\bar \rho}_x} \, .
\] 
{}From the last of equations \rf{string_eq4} one can obtain
expressions for the right and the left hand side 
of the above relation  in terms of ${\bar q}$ and $r$.
This results in a condition :
\begin{equation}
(\ln r)_x= -(\ln {\bar q})_x , \;\;\to \;\;{\bar q}=\frac{K}{r},
\;\;\; K={\rm const}\,,
\lab{KDB}
\end{equation}
which also easily  follows from the equality $w={\bar y}$.
The above relation resembles an effect of the
Darboux-B\"acklund transformation to be described in the next section.
Note that the above transformation maps $w$ to ${\bar y}$ and therefore
comparing with values of $\nu$ in eqs. \rf{painy} and \rf{painw}
we conclude that it lowers $\nu$ by $2$.

For completeness, let us mention that as follows from  eqs. \rf{px} for 
$Q=y$ and $P =-z/4y$, the function $Y=-2 P=z/2 y$ also satisfies the Painlev\'e 
IV equation \rf{painy} with the parameters 
${\widetilde \nu} = 3 \mu/2-\nu/2-2$, ${\widetilde \mu}= \pm(\mu+\nu)/2$. 
By comparing with definitions \rf{PQdefs} and definitions of $P_0,Q_0$
we find that the transformation $y \to Y$ is a superposition of the 
transformation $\rho \to {\widetilde \rho}$ discussed below equation \rf{jmo2} with the 
Darboux-B\"acklund transformation $y \to w$ lowering $\nu$ by $2$. 
In what follows we will only discuss the latter
transformation together with its square-root.

\section{Darboux-B\"acklund transformations}
The Darboux-B\"acklund transformation is here introduced 
as a similarity transformation with an operator $T=r \partial_x r^{-1}$
acting on the pseudo-differential AKNS Lax operator  through :
\[
L = \partial_x - r \partial_x^{-1} q \;\;\to \;\; {\bar L} = T\,L\,T^{-1}=
\partial_x - {\bar r} \partial_x^{-1} {\bar q}
\]
A simple calculation yields 
\begin{equation}
{\bar r} = r (\ln r)_{xx} - r^2 q , \quad {\bar q}=-\frac{1}{r}\, .
\lab{rq-db}
\end{equation}
in agreement with relation \rf{KDB} for $K=-1$.
Taking a product of ${\bar r}, {\bar q}$ one obtains
\[
{\bar \rho}_x = -2 {\bar r} {\bar q}= \rho_x +2 (\ln r)_{xx} \, .
\]
Thus the DB transformation \rf{rq-db} yields
\begin{align}
{\bar \rho} & = \rho +2 (\ln r)_{x}= \rho + 2 w +4 x =  \rho + 2 y_{-} +4 x\nonumber \\
&= \rho+2
\frac{-1}{2\rho_x}\left(2 \left(x \rho_x-\rho\right)-\rho_{xx}\right)+4 x 
=  \rho + 2 \frac{\rho}{\rho_x} + \frac{\rho_{xx}}{\rho_x}
+2x \lab{rho-DB}
\intertext{or including a reference to the $(\mu,\nu)$ parameters}
\rho^{(\mu,\nu-2)}&=\rho^{(\mu,\nu)} + 2 \frac{\rho^{(\mu,\nu)}}{\rho^{(\mu,\nu)}_x} + 
\frac{\rho^{(\mu,\nu)}_{xx}}{\rho^{(\mu,\nu)}_x} +2x \, .
\lab{rho-DB1}
\end{align}
It is equally easy to formulate the adjoint  Darboux-B\"acklund 
transformation which increases $\nu$ by $+2$.
The adjoint Darboux-B\"acklund transformation involves acting with
$S=q \partial_x q^{-1}$ on the pseudo-differential Lax operator
through :
\[
L = \partial_x - r \partial_x^{-1} q \;\,\to \;\, {\widetilde L} = S^{*\,-1}\,L\,
S^{*}= (q^{-1} \partial_x^{-1} q) L (q^{-1} \partial_x q)=
\partial_x - {\widetilde  r} \partial_x^{-1} {\widetilde q}
\]
with
\begin{equation}
{\widetilde q} = -q (\ln q)_{xx} + q^2 r , \quad {\widetilde r}=\frac{1}{q}\, .
\lab{rq-adb}
\end{equation}
Taking a product of ${\widetilde  r}, {\widetilde q}$ one obtains
\[
{\widetilde \rho}_x = -2 {\widetilde  r} {\widetilde q}= \rho_x +2 (\ln q)_{xx}
\]
Thus the DB transformation \rf{rq-adb} yields
\begin{align}
{\widetilde \rho} & = \rho +2 (\ln q)_{x}= \rho -2 y -4 x =  
\rho - 2 y_{+} -4 x \nonumber\\
&=  \rho-2\frac{-1}{2\rho_x}\left(2 \left(x \rho_x-\rho\right)+\rho_{xx}\right)-4 x 
=  \rho - 2 \frac{\rho}{\rho_x} + \frac{\rho_{xx}}{\rho_x}
-2x \lab{rho-aDB}
\intertext{or}
\rho^{(\mu,\nu+2)}&=\rho^{(\mu,\nu)} - 2 \frac{\rho^{(\mu,\nu)}}{\rho^{(\mu,\nu)}_x} + 
\frac{\rho^{(\mu,\nu)}_{xx}}{\rho^{(\mu,\nu)}_x} -2x \, .
\lab{rho-aDB1}
\end{align}
Transformations ${^{\sim}}$ and $\bar{\phantom{ a}}$
when acting on variables $J,\bj$ such that:
\[
\bj= -rq =  \rho_x/2, \quad J = (\ln q)_x=-y-2x
\]
take the following form \cite{Aratyn:1993zi} :
\begin{xalignat}{2}
G (J)  & \equiv  J + \left( \ln \left( \bj + J_x \right) \right)_x &
G ( \bj)& \equiv \bj + J_x
\lab{jtransf} \\
G^{-1} (J) &\equiv J - \left( \ln \bj \right)_x  &
 G^{-1} (\bj) &\equiv
\bj + \left( \ln \bj  \right)_{xx} - J_x\, ,
\lab{jtransf2}
\end{xalignat}
where we found it convenient to rewrite actions of ${^{\sim}}$ and $\bar{\phantom{ a}}$
as transformations $G$ and $G^{-1}$.
It follows from \rf{jtransf}-\rf{jtransf2} that:
\begin{equation}
G(y) = y- \left(\ln \left( y_x+y^2+2xy+2 \nu +4\right)\right)_x, \;
G^{-1}(y) = y+ \left(\ln \left( y_x-y^2-2xy-2 \nu\right)\right)_x\, .
\lab{Grho}
\end{equation}
The DB transformations $G^{\pm1}$ are equal to Murata's transformations
$T_{\mp}$ first obtained in \cite{murata}. This is established after making
use of the Painlev\'e  equation \rf{painy} in \rf{Grho} and identifying
parameters $\theta, \alpha$ from \cite{murata} with $\mu, -\nu-1$.

There also exists a set of variables $\sj,\bsj$ 
related to $J,\bj$ via a 
Miura transformation \cite{Aratyn:1993zi} such that
\begin{equation}
J = - \sj \,- \bsj \, + \frac{\sj_x}{\sj}  \qquad ; \qquad
\bj = \bsj \, \sj            \lab{miura1}
\end{equation}
In terms of variables  $\sj,\bsj$ one can define a discrete
symmetry transformation 
\begin{xalignat}{2}
g (\sj\,)  & \equiv  \bsj\, - \frac{\sj_x}{\sj} &
g\;( \bsj\,) &\equiv \sj
\lab{sjtransf} \\
g^{-1} (\bsj\,) &\equiv \sj\, + \frac{ \bsj_x}{ \bsj  }
&g^{-1} ( \sj\,)& \equiv \bsj
\lab{sjtransf2}
\end{xalignat}
such that \cite{Aratyn:1993zi} 
\begin{equation} g^2=G\, ,
\lab{gsquare}
\end{equation}
when applied on $J, \bj$ defined by equation \rf{miura1}.

{}From the above relations one obtains simple transformation rules:
\begin{equation}
g(y)=y-\left(\ln (-\sj+y+2x)\right)_x, \;\;\;\,
g^{-1}(y)=y+\left(\ln (\sj)\right)_x\,.
\lab{ggy}
\end{equation}
Acting on solutions $r,q$ or $J,\bj$ of the Painlev\'e IV equation
with $G$ raises $\nu$ by $2$ and keeps $\mu$ constant while $g$ 
when applied on solutions expressed by $\sj,\bsj$ shifts 
both $\nu$ and $\mu$ by $1$ (see below) in such a way that acting
twice with $g$ agrees with the formula \rf{gsquare}.
Because of property of the $g$ transformation to shift both parameters 
of the Painlev\'e equation this transformation will be very useful
in what follows in deriving solutions corresponding to new values of 
the parameters.

Plugging $r,q$ into expression \rf{miura1} leads to 
\begin{equation}
(\ln q )_x= -\bsj-\sj+ \sj_x/\sj, \quad\,\;
rq=-\frac12 \rho_x =-\bsj \sj  \, ,
\lab{miura2}
\end{equation}
which after elimination of  $\bsj$  yields a Riccati 
equation for $\sj$ :
\begin{equation}
rq = \sj^2+\sj (\ln q )_x - \sj_x\, .
\lab{riccatij}
\end{equation}
Rewriting equation \rf{px} as $ 2 QP+\left( \mu +\nu\right)
=(2 P)^2 +2P(-Q -2 x)- 2P_x$ and recalling that
$Q=-(\ln q )_x-2x$ and $2 QP+\left( \mu +\nu\right)=rq$ we find
after comparing with relation \rf{riccatij} that the solution 
to this relation is given by
\begin{equation}
j= 2P = -\frac{z}{2y}=\frac{1}{2y} \left(y^2-y_x+2xy-2 \mu\right)\, ,
\lab{jtwop}
\end{equation}
where use was made of the first equation in \rf{pain1}.
Plugging the value of $j$ from \rf{jtwop} with two values of 
$\mu= \pm \vert \mu\vert$
into  relations \rf{ggy} one obtains two set of transformations 
$g_{\pm}:\; \nu \to \nu+1, \, \mu^2 \to (\vert \mu\vert \pm 1)^2$
and two set of inverse transformations
$g^{-1}_{\pm}:\; \nu \to \nu-1, \, \mu^2 \to (\vert \mu\vert \pm 1)^2$
recovering expressions found in \cite{fokas}.

As observed in \cite{Aratyn:1993zi}, if one sets
$S_n = G^n \left( J\right)$ and $R_n = G^n \left( \bj \right)$ 
then the above  set of variables satisfy
the equations of motion of the Toda chain:
\[ 
\partial_x \,S_n =  R_{n +1} - R_n  , \quad  
\partial_x\, R_n = R_n \left( S_{n} - S_{n-1} \right) \, .
\]
For quantities
$\p_n \equiv \int G^n \left( J\right) dt$ and 
$\partial_x \ln \tau_n \equiv  \int G^n \left( \bj \right) dt$
one finds on basis of properties \rf{jtransf} and \rf{jtransf2} 
of the $G$ symmetry :
\begin{equation}
\frac{\tau_{n+1}}{ \tau_{n}} \, = \, e^{\p_n} \qquad;\qquad
e^{\p_n - \p_{n-1}} \, = \,  G^n \left( \bj \right) \, =\, R_n  \, ,        \lab{taueqs}
\end{equation}
from which the Hirota type of Toda chain equation:
\begin{equation}
\partial^2_x \ln \tau_n  = \frac{\tau_{n+1} \, \tau_{n-1} }{ \tau^2_{n}}
\lab{hirotaeq}
\end{equation}
follows easily. The symmetry structure of $g$ transformations is that of 
Volterra lattice \cite{Aratyn:1993zi}.

\subsection{ Darboux-B\"acklund transformations and ``$-2x$ hierarchy''}
Things simplify in the case of $q=0, r=r^{(0)}$ and 
thus with  the initial  AKNS Lax operator $L^{(0)}= \partial_x $.
A chain of Darboux-B\"acklund transformations
define a string of new Lax operators via:
\[
L^{(0)}= \partial_x  \; \longrightarrow\; L^{(k)}= \partial_x +r^{(k)} \partial_x ^{-1}q^{(k)}
\]
with
\[
 L^{(k)}= T^{(k)}\left( r^{(k-1)}\right) \partial_x \left(T^{(k)}
 \left( r^{(k-1)}\right)\right)^{-1} , \;\;\, T^{(k)}\left( r^{(k-1)}\right) =
 r^{(k-1)}  \partial_x \left( r^{(k-1)}\right)^{-1}  
\]
for $k \ge 1$. 
The eigenfunction $r^{(0)}$ of $L^{(0)}$ has to satisfy
$\partial_{t_n}r^{(0)}= \left(L^{(0)}\right)^n_{+} r^{(0)} = \partial_x^n r^{(0)}$.
The standard choice 
\[
r^{(0)} (t)= e^{\sum_1^\infty t_n z^n} = \sum_{k=0}^\infty S_k (t) z^k
\]
ensures that this is the case.
To be able to assign a simple conformal weight to $r$ one can break this 
sum and just set
$r^{(0)} (t)= S_n (t)$ due to the scaling law :
\[ S_n (t_\lambda ) = \lambda^n  S_n (t ) , \quad t_\lambda = (\lambda x, \lambda^2 t_2, {\ldots} )
\]
which can be expressed as a differential condition (see 
\rf{string_eq3}):
\[
\left(x \partial_x +2 t_2 \pder{}{t_2} + {\ldots} \right)  S_n (t )=
n S_n (t ),\quad \to \quad n=\nu-1\, .
\]
With this choice the Wronskian expression from  \cite{Aratyn:1996nb} becomes :
\[
r^{(k)} =r^{(k)}= \frac{W_{k+1} \lbrack r^{(0)}, \partial_x r^{(0)}, {\ldots} , \partial_x^k r^{(0)}\rbrack}
{W_{k} \lbrack r^{(0)}, \partial_x r^{(0)}, {\ldots} , \partial_x^{k-1} r^{(0)}\rbrack}
= \frac{W_{k+1} \lbrack S_n, S_{n-1}, {\ldots} , S_{n-k}\rbrack}
{W_{k} \lbrack S_n, S_{n-1}, {\ldots} ,  S_{n-k+1}\rbrack}\, ,
\]
after using $S_{n-k}=\partial_x^k S_n$.

Similarly we find for $q^{(k)}$
\[
q^{(k)}=-\frac{1}{r^{(k-1)}}= -\frac{W_{k-1} \lbrack S_n,  S_{n-1}, 
{\ldots} , S_{n-k+2}\rbrack}
{W_{k} \lbrack S_n, S_{n-1}, {\ldots} , S_{n-k+1}\rbrack}, \;\;\;
k>1\, .
\]
We see that the above Darboux-B\"acklund transformation
changes the value of $\nu$ by increasing the value of $k$.

Note that for $t_2=-1/4$ and $t_3=t_4={\ldots} =0$
one gets
\[ e^{\sum_1^\infty t_n z^n}=e^{xz-z^2/4}=
\sum_{n=0}^\infty \frac{1}{n!} H_n (x) \left(\frac{z}{2}\right)^n
\]
and so in this limit $S_n (x) = H_n(x) / 2^n n!$.
Thus $r^{(0)}=H_n(x)$ and $q^{(0)}=0$ and $\nu =n+1$
agrees with what we found above for solution with 
$q=0$ and $\nu$ being a positive integer.

Note that a simple Wronskian formula gives:
\[
\begin{split}
r^{(k)} q^{(k)}&=
-\frac{W_{k+1} \lbrack S_n, S_{n-1}, {\ldots} , S_{n-k}\rbrack\,
W_{k-1} \lbrack S_n,  S_{n-1}, 
{\ldots} , S_{n-k+2}\rbrack}
{W_{k}^2 \lbrack S_n, S_{n-1}, {\ldots} ,  S_{n-k+1}\rbrack}\\
&= -\partial_x^2 \ln W_{k} \lbrack S_n, S_{n-1}, {\ldots} , S_{n-k+1}\rbrack
\end{split}
\]
and so one finds
\[
\rho^{(k,n)} = 2 \partial_x \ln W_{k} \lbrack S_n, S_{n-1}, {\ldots} , S_{n-k+1}\rbrack
\]
or in terms of the Hermite polynomials:
\[
\rho^{(k,n)} = 2 \partial_x \ln W_{k} \lbrack H_n, H_{n-1}, {\ldots} ,
H_{n-k+1}\rbrack \,, \;\;\; k>1\, ,
\]
which satisfies $\rho$-equation \rf{rho-eq} with $\nu
=n-2\cdot k+1$ and $\mu^2=(n+1)^2$. That result is in 
agreement with 
\[
w^{(k,n)}=\left(\ln r^{(k)}\right)_x-2x=
\partial_x \left(\ln \frac{W_{k+1} \lbrack H_n, H_{n-1}, {\ldots} , H_{n-k}\rbrack}
{W_{k} \lbrack H_n,  H_{n-1}{\ldots} , H_{n-k+1}\rbrack} \right)
-2x
\]
satisfying \rf{painw} with the same values of parameters.

There exists another chain of successive DB transformations,
which also builds the AKNS Lax pseudo-differential operator
starting from the ``pure'' derivative $\partial_x$
by :
\[
L^{(0)}= \partial_x  \; \longrightarrow\; L^{(k)}= \partial_x +r^{(k)} \partial_x ^{-1}q^{(k)}
\]
with
\[
 L^{(k)}= T^{(k)}\left( q^{(k-1)}\right) \partial_x \left(T^{(k)}
\left( q^{(k-1)}\right)\right)^{-1} , \;\;\,  T^{(k)}\left( q^{(k-1)}\right) =
 \left( q^{(k-1)}\right)^{-1}  \partial_x^{-1} q^{(k-1)}  
\]
for $k \ge 1$.
The adjoint eigenfunction $q^{(0)}$ of $L^{(0)}$ has to satisfy
$\partial_{t_n}q^{(0)}= -\left(L^{(0)}\right)^{*\,n}_{+} q^{(0)} = 
(-1)^{n+1} \partial_x^n q^{(0)}$.
The standard choice 
\[
q^{(0)} (t)= e^{\sum_1^\infty (-1)^{n+1}t_n z^n} = 
\sum_{k=0}^\infty {\widehat S}_k (t) z^k
\]
ensures that this is the case.
To be able to assign a simple conformal weight for $q$ it is convenient
to break this sum by setting
$q^{(0)} (t)= {\widehat S}_n (t)$ due to the scaling law :
\[ {\widehat S}_n (t_\lambda ) = \lambda^n  {\widehat S}_n (t ) , \quad t_\lambda = (\lambda x, \lambda^2 t_2, {\ldots} )
\]
With this choice the Wronskian expression for $q^{(k)}$   becomes
\[
q^{(k)}=- \frac{W_{k+1} \lbrack q^{(0)}, \partial_x q^{(0)}, {\ldots} , \partial_x^k q^{(0)}\rbrack}
{W_{k} \lbrack q^{(0)}, \partial_x q^{(0)}, {\ldots} , \partial_x^{k-1} q^{(0)}\rbrack}
 = -\frac{W_{k+1} \lbrack {\widehat S}_n, {\widehat S}_{n-1}, {\ldots} , {\widehat S}_{n-k}\rbrack}
{W_{k} \lbrack {\widehat S}_n, {\widehat S}_{n-1}, {\ldots} ,  {\widehat
S}_{n-k+1}\rbrack}  
\]
after using ${\widehat S}_{n-k}=\partial_x^k {\widehat S}_n$.
Similarly we find for $r^{(k)}$
\[
r^{(k)}=r^{(k)}=\frac{1}{q^{(k-1)}}= -\frac{W_{k-1} \lbrack {\widehat S}_n,  {\widehat S}_{n-1}, 
{\ldots} , {\widehat S}_{n-k+2}\rbrack}
{W_{k} \lbrack {\widehat S}_n, {\widehat S}_{n-1}, {\ldots} , {\widehat S}_{n-k+1}\rbrack}, \;\;\;
k>1\, .
\]
Note that for $t_2=-1/4$ and $t_3=t_4={\ldots} =0$
one gets
\[ e^{\sum_1^\infty t_n (-1)^{n+1}z^n}=e^{xz+z^2/4}=
\sum_{n=0}^\infty \frac{1}{n!} {\widehat H}_n (x) \left(\frac{z}{2}\right)^n
\]
and so in this limit ${\widehat S}_n (x) = {\widehat H}_n(x) / 2^n n!$,
where ${\widehat H}_n(x)$ is introduced in eq. \rf{hathermit}
Thus $q^{(0)}={\widehat H}_n(x)$ and $r^{(0)}=0$ 
and $\nu =-n-1$
agrees with what we found above for solution with 
$r=0$ and $\nu<0$ and integer.
Indeed 
\[
{\widehat \rho}^{(k,n)} = 2 \partial_x \ln W_{k} \lbrack {\widehat H}_n, {\widehat H}_{n-1},{\ldots} , 
{\widehat H}_{n-k+1}\rbrack
\]
satisfies $\rho$-equation \rf{rho-eq} with $\nu
=-n+2\cdot k-1$ and $\mu^2=(n+1)^2$.

To summarize we found the solutions to the $\rho$-equation \rf{rho-eq} with 
$\nu, \mu$ parameters given by :
\begin{equation}
\begin{split}
\rho^{(k,n)} &= 2 \partial_x \ln W_{k} \lbrack {H}_n, { H}_{n-1},{\ldots} , 
{ H}_{n-k+1}\rbrack, \quad \nu =n-2\cdot k+1, \;\;\mu^2=(n+1)^2\\
{\widehat \rho}^{(k,n)}&= 2 \partial_x \ln W_{k} \lbrack {\widehat H}_n, {\widehat H}_{n-1},{\ldots} , 
{\widehat H}_{n-k+1}\rbrack, \quad \nu=-n+2\cdot k-1, \;\;
\mu^2=(n+1)^2
\end{split}
\lab{rhos}
\end{equation}
with $n,k=1,2,3,{\ldots} $.
Corresponding solutions to the Painlev\'e eqs. \rf{painy} and \rf{painw},
referred to as ``$-2x$-hierarchy''\cite{Clarkson-jmp,Clarkson}
are, respectively,
\begin{equation}
\begin{split}
w^{(k,n)} &= \partial_x \ln r^{(k-1)} -2 x =
\partial_x \ln \frac{W_{k} \lbrack {H}_n, { H}_{n-1},{\ldots} , 
{ H}_{n-k+1}\rbrack}{W_{k-1} \lbrack {H}_n, { H}_{n-1},{\ldots} , 
{ H}_{n-k+2}\rbrack} -2 x \\
\nu &=n-2\cdot k+1, \;\;\mu^2=(n+1)^2
\end{split}
\lab{wyfr}
\end{equation}
\begin{equation}
\begin{split}
y^{(k,n)} &=  -\partial_x \ln q^{(k-1)} -2 x=
-\partial_x \ln \frac{W_{k} \lbrack {\widehat H}_n, {\widehat H}_{n-1},{\ldots} , 
{\widehat H}_{n-k+1}\rbrack}{W_{k-1} \lbrack {\widehat H}_n, {\widehat H}_{n-1},{\ldots} , 
{\widehat H}_{n-k+2}\rbrack} -2 x\\
\nu&=-n+2\cdot k-1, \;\;
\mu^2=(n+1)^2
\end{split}
\lab{wyfq}
\end{equation}
with $n,k=1,2,3,{\ldots} $.

\subsection{Darboux-B\"acklund transformations and ``$-1/x$-hierarchy''}
We will now examine an hierarchy of solutions which can be obtained by the
Darboux-B\"acklund approach from remaining basic solutions 
listed in eqs. \rf{qhermit} and  \rf{rhermit}. Namely, we consider as 
starting points $ q(x) =  \exp \left(-x^2\right)  H_{\nu} (x)$ with 
$\nu$ being a positive integer
and $r(x) =   e^{x^2} {\widehat H}_{-\nu} ( x)$
for $\nu \le 0$ and integer.

We begin with $ q^{(0)}=  \exp \left(-x^2\right)  H_{m} (x)$, 
$r^{(0)}=0$ and
corresponding
\[y =- \partial_x \ln \left(e^{-x^2}  H_{m}\right) -2 x =- \partial_x \ln  H_{m}, 
\]
which satisfies eq. \rf{painy} with $\nu=m$ and $\mu^2=m^2$.

The iteration procedure \rf{jtransf} expressed here in terms
of the AKNS variables :
\begin{equation}
\begin{split}
q^{(1)}&=G(q^{(0)})= - q^{(0)} \left(\ln q^{(0)}\right)_{xx}+ (q^{(0)})^2 r^{(0)}, \\
q^{(k+1)}&= G(q^{(k)})= - q^{(k)} \left(\ln q^{(k)}\right)_{xx} + \left(q^{(k)}\right)^2
\frac{1}{q^{(k-1)}} ,\quad k >0
\end{split}
\lab{qtransf}
\end{equation}
with $q^{(0)},r^{(0)}$ as defined above is solved by
\begin{equation}
q^{(k)}= \frac{W_{k+1} \lbrack e^{-x^2} H_m, e^{-x^2} H_{m+1}, 
{\ldots} , e^{-x^2} H_{m+k} \rbrack}
{W_{k} \lbrack  e^{-x^2} H_m, e^{-x^2} H_{m+1}, 
{\ldots} , e^{-x^2} H_{m+k-1}  \rbrack} \, .
\lab{qke}
\end{equation}
Using the recurrence relation 
$ \left( \exp (-x^2) H_m \right)_x = -\exp (-x^2) H_{m+1}$
for Hermite's polynomials allows us to cast the above 
expression for $q^{(k)}$ into a simple form
\[
q^{(k)}= (-1)^k \,\frac{W_{k+1}}{W_{k}} 
\]
where 
\[ W_{k+1} = W_{k+1} \lbrack f, f_x, f_{xx}, {\ldots} ,
f_{\underbrace{x\cdots x}_{k}}\rbrack
, \quad f=  e^{-x^2} H_m
\]
Then applying the $G$ transformation yields:
\[
\begin{split}
G(q^{(k)})
&= (-1)^k \frac{W_{k}}{W_{k+1}} \left[  \frac{1}{W_{k}^2}\left( - W_{k+1}
W_{k+1}^{\prime\prime}+ (W_{k+1}^{\prime})^2 \right) +
\frac{W_{k+1}^2W_{k}^{\prime\prime}}{W_{k}^{3}} \right.\\
&\left. - \frac{W_{k+1}^2 (W_{k}^{\prime})^2 }{W_{k}^{4}}
-\frac{W_{k+1}^3 W_{k-1}}{W_{k}^4}\right] \\
&= (-1)^{k+1}\frac{W_k^2 W_{k+2}}{W_{k+1}W_k^2}
=(-1)^{k+1}\frac{W_{k+2}}{W_{k+1}}
\end{split}
\]
where use was repeatedly made of the identity
$
W_k W_k^{\prime\prime}- (W_{k}^{\prime})^2= W_{k-1}W_{k+1}
$ 
for an arbitrary argument $f$.
Thus, $G(q^{(k)}) = q^{(k+1)}$
and \rf{qke} is established by an induction argument.

Taking into account the identity
\[
W_{k} \lbrack e^{-x^2} P_1,{\ldots} ,  e^{-x^2} P_k
\rbrack= W_{k} \lbrack P_1,{\ldots} ,   P_k
\rbrack\; e^{-k\, x^2}
\]
and setting $m=n-k+1$ we obtain 
the following solution to the Painlev\'e eq. \rf{painy}:
\begin{equation}
\begin{split}
y^{(k,n)}
&= -\partial_x 
\ln \frac{W_{k+1} \lbrack e^{-x^2} {H}_{n+1}, e^{-x^2} { H}_{n},
{\ldots}, e^{-x^2} { H}_{n-k+1}\rbrack}{W_{k} \lbrack e^{-x^2} {H}_{n},e^{-x^2} { H}_{n-1},
{\ldots} , e^{-x^2} { H}_{n-k+1}\rbrack}- 2x\\
&=  \partial_x \ln \frac{W_{k} \lbrack {H}_{n}, { H}_{n-1},
{\ldots} , { H}_{n-k+1}\rbrack}{W_{k+1} \lbrack {H}_{n+1}, { H}_{n},{\ldots} , 
{ H}_{n-k+1}\rbrack}
\end{split} 
\lab{wiiq}
\end{equation}
with $\nu= n+k+1$ and $\mu^2= (n-k+1)^2$.
Note that $y^{(0,n)}$ reproduces
solution $-\partial_x \ln { H}_{n}$ encountered above.

We now turn our attention to $r(x) =r^{(0)} (x) =   e^{x^2} {\widehat H}_{m} ( x)$
and corresponding $w=\partial_x \ln \left({\widehat H}_{m}\right)$,
which satisfies \rf{painw}
with $\nu=-m$ and $\mu^2=m^2$  in agreement with eq. \rf{rhermit}.
Repeating steps applied above to find $q^{(k)}$ one easily establishes that
\begin{equation}
r^{(-k)}= \frac{W_{k+1} \lbrack e^{x^2} {\widehat H}_{m}, 
e^{x^2} {\widehat H}_{m+1}, 
{\ldots} , e^{x^2} {\widehat H}_{m+k} \rbrack}
{W_{k} \lbrack  e^{x^2} {\widehat H}_{m}, e^{x^2} {\widehat H}_{m+1}, 
{\ldots} , e^{x^2} {\widehat H}_{m+k-1}  \rbrack} \, .
\lab{rke}
\end{equation}
happens to be a solution to the iteration relations:
\begin{equation}
\begin{split}
r^{(-1)}&=G^{-1} (r^{(0)})= r^{(0)} \left(\ln r^{(0)}\right)_{xx} -
(r^{(0)})^2 q^{(0)},\\
r^{(-k-1)}&= G^{-1} (r^{(-k)})= r^{(-k)} \left(\ln r^{(-k)}\right)_{xx} + \left(r^{(-k)}\right)^2
\frac{1}{r^{(-k+1)}} ,\quad k >0\,.
\end{split}
\lab{rtransf}
\end{equation}
with $q^{(0)}=0$.
Plugging expression \rf{rke} into expression for $w$
from relation \rf{ywz} and setting $m=n-k+1$
one obtains solutions
\begin{equation}
w^{(k,n)}= - \partial_x \ln \frac{W_{k} \lbrack {\widehat H}_{n}, { \widehat H}_{n-1},
{\ldots} , { \widehat H}_{n-k+1}\rbrack}{W_{k+1} \lbrack {\widehat H}_{n+1}, { \widehat H}_{n},{\ldots} , 
{ \widehat H}_{n-k+1}\rbrack}
= \partial_x \ln \frac{W_{n-k+1} \lbrack {H}_{n+1}, {H}_{n},
{\ldots} , { H}_{k+1}\rbrack}{W_{n-k+1} \lbrack {H}_{n}, {H}_{n-1},{\ldots} , 
{H}_{k}\rbrack}
\lab{hwiiiq}
\end{equation}
to the Painlev\'e eq. \rf{painw} with $\nu=-(n+k+1), \; \mu =(n-k+1)^2$.
In obtaining \rf{hwiiiq} use was made of a Wronskian identity,
\[
\frac{W_{k+1} \lbrack {\widehat H}_{n+1}, { \widehat H}_{n},{\ldots} , 
{ \widehat H}_{n-k+1}\rbrack}{W_{k} \lbrack {\widehat H}_{n}, { \widehat H}_{n-1},
{\ldots} , { \widehat H}_{n-k+1}\rbrack}= C_{n,k}
\frac{W_{n-k+1} \lbrack {H}_{n+1}, {H}_{n},
{\ldots} , { H}_{k+1}\rbrack}{W_{n-k+1} \lbrack {H}_{n}, {H}_{n-1},{\ldots} , 
{H}_{k}\rbrack}\, ,
\]
where $C_{n,k}$ are some combinatorial constants.

Expressions \rf{wiiq} and \rf{hwiiiq} define ``$-1/x$-hierarchy''
\cite{Clarkson-jmp}.

\subsection{Darboux-B\"acklund transformations and ``$-2x/3$-hierarchy''}
\subsubsection{Solutions with $\mathbf{\mu^2=(1/3)^2, (2/3)^2}$}

Consider now solution \rf{rhob1} of the  $\rho$-equation \rf{rho-eq}
with $\mu^2=(2/3)^2,\nu=0$.
According to relation \rf{yrho} the corresponding two solutions of 
equations \rf{painy}, \rf{painw}
are :
\[
y_{+}= -\frac{1}{x} \left( \frac{2}{3} x^2 +1\right)=-(\ln q)_x-2x,
\quad y_{-}= -\frac{1}{x} \left( \frac{2}{3} x^2 -1\right)=(\ln r)_x-2x,
\]
with the following solutions of the string equations \rf{string_eq4} 
for $\nu=0$:
\begin{equation}
q =q^{(0)}= -\frac{2}{3} x e^{-2x^2/3} 
=F_0^{(1)} e^{-2x^2/3} , \;\;\;
r=r^{(0)}=  \frac{2}{3} x e^{2x^2/3} 
= {\widehat F}_0^{(1)}  e^{2x^2/3} \, .
\lab{rqrhob1}
\end{equation}
In the above equation we employed notation involving polynomials :
\begin{equation}
F_n^{(k)}=  \frac{e^{x^2/3}}{2^n n!} \dder[3n+k]{}{x} e^{-x^2/3},\;\;
{\widehat F}_n^{(k)}= \, \frac{ e^{-x^2/3}}{2^n n!} \dder[3n+k]{}{x} e^{x^2/3},
\lab{fnk}
\end{equation}
defined for  $k=0,1,2, \;\;n=0,1,2,{\ldots}$. By simple rescaling of their
arguments polynomials $F_n^{(k)}, {\widehat F}_n^{(k)}$ become proportional to
Hermite polynomials $H_m , {\widehat H}_m$ for certain values
of $m$.

\begin{subequations}\label{eq:twothird}
Acting on the initial configuration \rf{rqrhob1} successively with 
the DB transformations as in \rf{qtransf} one arrives at the following 
Wronskian representation in terms of the ratios of Wronskians :
\begin{align}
q^{(n)}&=\frac{W_{n+1} \lbrack 
F_0^{(1)}, F_1^{(1)}, {\ldots} , F_n^{(1)}\rbrack}
{W_{n} \lbrack F_0^{(1)}, F_1^{(1)}, {\ldots} , F_{n-1}^{(1)}\rbrack}
\exp \left( -\frac{2 x^2}{3}\right),
\quad r^{(n)}=1/q^{(n-1)} \lab{qtwothird} \\
\mu&=\pm \frac{2}{3} , \;\; \nu =2n ,\;\; n \ge 0\, .\nonumber
\end{align}
Acting with negative powers of $G$ as in \rf{rtransf}  yields :
\begin{align}
 r^{(-n)}&=\frac{W_{n+1} \lbrack 
 {\widehat F}_0^{(1)}, {\widehat F}_1^{(1)}, {\ldots} , {\widehat F}_n^{(1)}\rbrack}
 {W_{n} \lbrack {\widehat F}_0^{(1)}, {\widehat F}_1^{(1)}, {\ldots} , {\widehat F}_{n-1}^{(1)}\rbrack}
 \exp \left( \frac{2x^2}{3}\right),
 \quad q^{(-n)}=-1/r^{(-n+1)} \lab{rtwothird} \\
 \mu&=\pm \frac{2}{3} , \;\; \nu =-2n ,\;\; n \ge 0 \, .\nonumber
\end{align}
\end{subequations}
The first few solutions are
\begin{xalignat*}{2}
q^{(1)}&=  \frac{2\left(-12x^2+4x^4-9\right)}{27x}
e^{- \tfrac{2x^2}{3}}, &
q^{(2)}&= - \frac{8x(504x^4-192x^6+16x^8-2835)}
{243(-12x^2+4x^4-9)}e^{- \tfrac{2x^2}{3}}\\
r^{(-1)} &= 
\frac {2(12\,x^{2} - 9 + 4\,x^{4})}{27x}
e^{ \tfrac{2x^2}{3}}, &
r^{(-2)} &= \frac{8x\,(- 2835 + 504\,x^{4} + 192\,x^{6} + 16
 \,x^{8})}{243(12\,x^{2} - 9 + 4\,x^{4})}e^{\tfrac{2x^2}{3}}\,.
\end{xalignat*}

Applying recursively \rf{rho-DB1} one gets closed
Wronskian expressions for the $\rho$-function:
\[ 
\begin{split}
\rho^{(n)}&=G^{n} (\rho)=
\rho - 2n \frac{4x}{3}+  
2 \left(\ln W_{n} \lbrack F_0^{(1)} , F_1^{(1)}, {\ldots} , 
F_{n-1}^{(1)}\rbrack\right)_x,  \\
\rho^{(-n)}&=G^{-n} (\rho)=
\rho + 2n \frac{4x}{3}+  
2 \left(\ln W_{n} \lbrack {\widehat F}_0^{(1)} , {\widehat F}_1^{(1)}, {\ldots} , 
{\widehat F}_{n-1}^{(1)}\rbrack\right)_x,  
\end{split}
\] 
which are solutions of the $\rho$-equation \rf{rho-eq}
with $\nu=\pm 2n, n=1,2, 3, {\ldots}$, respectively.
In particular 
\[
\rho^{(\pm 1)}= \frac{2}{27x} \left(4x^4+27\pm36x^2\right)
\]
satisfies the $\rho$-equation \rf{rho-eq} with $\mu^2 = \frac49,
\nu = \pm 2$.

Consider now the basic solutions \rf{rho13} of
$\rho$-equation \rf{rho-eq}.
Plugging \rf{rho13} (with a plus sign) into relation \rf{yrho} one obtains the following
two solutions of equation \rf{painy}:
\begin{xalignat}{2}
y=y_{+}&= - \frac23 x, &  \mu &= \pm \frac13, \;\; \nu = -1 \\
w=y_{-}&= - \frac23 x \frac{\frac{16}{9} x^2 -\frac{8}{3}}
{\frac{16}{9} x^2 +\frac{8}{3}}
, &  \mu &= \pm \frac13, \;\;\; \nu = -1 
\label{eq:ypm}
\end{xalignat}
Eq. \rf{ypm} corresponds to the following AKNS variables:
\begin{equation}
\begin{split}
r &= r^{(0)}=\frac23 \left(  \frac23 x^2+1\right)e^{2x^2/3}=
{\widehat F}_0^{(2)}  e^{2x^2/3},\\ 
q&= q^{(0)}=  -  e^{-2x^2/3}=-F_0^{(0)} e^{-2x^2/3},
\end{split}
\lab{rqrho13}
\end{equation}
\begin{subequations}\label{eq:monethird}
Applying the DB transformations $G,G^{-1}$ generalizes the above basic
solutions to 
\begin{align}
q^{(n)}&=-\frac{W_{n+1} \lbrack 
F_0^{(0)}, F_1^{(0)}, {\ldots} , F_n^{(0)}\rbrack}
{W_{n} \lbrack F_0^{(0)}, F_1^{(0)}, {\ldots} , F_{n-1}^{(0)}\rbrack}
\exp \left( -\frac{2x^2}{3}\right),
\quad r^{(n+1)}=1/q^{(n)}  \lab{qmonethird}\\
\mu&=\pm \frac{1}{3} , \;\; \nu =-1+2n ,\;\;\; n \ge 0  \nonumber
\end{align}
and
\begin{align}
r^{(-n)}&=\frac{W_{n+1} \lbrack 
{\widehat F}_0^{(2)}, {\widehat F}_1^{(2)}, {\ldots} , {\widehat F}_n^{(2)}\rbrack}
{W_{n} \lbrack {\widehat F}_0^{(2)}, {\widehat F}_1^{(2)}, {\ldots} , {\widehat F}_{n-1}^{(2)}\rbrack}
\exp \left( \frac{2x^2}{3}\right),
\quad q^{(-n)}=-1/r^{(-n+1)} \lab{rmonethird}\\
\mu&=\pm \frac{1}{3} , \;\;\;  \nu =-1-2n ,\;\;\; n \ge 0\, . \nonumber
\end{align}
\end{subequations}
Moreover one obtains the following Wronskian expressions for the chain
of associated solutions to the $\rho$-equation:
\[ 
\begin{split}
\rho^{(n)}&=G^{n} (\rho)=
\rho - 2n \frac{4x}{3}+  
2 \left(\ln W_{n} \lbrack  F_0^{(0)}, F_1^{(0)}, {\ldots} , F_{n-1}^{(0)}
\rbrack\right)_x,  \;\;\nu=1+2n\, , \\
\rho^{(-n)}&=G^{-n} (\rho)=
\rho + 2n \frac{4x}{3}+  
2 \left(\ln W_{n} \lbrack  {\widehat F}_0^{(2)}, {\widehat F}_1^{(2)}, {\ldots} , {\widehat F}_{n-1}^{(2)}
\rbrack\right)_x,  \;\;\nu=1-2n
\end{split}
\] 
with $n=1,2,3,{\ldots} $.

Plugging \rf{rho13} (with a minus sign) into relation \rf{yrho} one obtains the following
two solutions of equation \rf{painy}:
\begin{xalignat}{2}
y=y_{+}&= - \frac23 x \frac{3+2x^2}{-3+2x^2}
, &  \mu &= \pm \frac13, \;\; \nu = +1 \\
w=y_{-}&= - \frac23 x 
, &  \mu &= \pm \frac13, \;\; \nu = +1 
\label{eq:ypp}
\end{xalignat}
which corresponds to the following AKNS variables:
\[
r=r^{(0)}= {\widehat F}_0^{(0)}  e^{2x^2/3},\;\; q=q^{(0)}= - F_0^{(2)} e^{-2x^2/3},\;\;
\mu=\pm \frac{1}{3}, \nu=1 \, .
\]
\begin{subequations}\label{eq:ponethird}
Applying the DB transformations   $G,G^{-1}$ generalizes the above basic
solutions to 
\begin{align}
q^{(n)}&=\frac{W_{n+1} \lbrack 
F_0^{(2)}, F_1^{(2)}, {\ldots} , F_n^{(2)}\rbrack}
{W_{n} \lbrack F_0^{(2)}, F_1^{(2)}, {\ldots} , F_{n-1}^{(2)}\rbrack}
\exp \left( -\frac{2x^2}{3}\right),
\quad r^{(n)}=1/q^{(n-1)} \lab{qponethird} \\
\mu&=\pm \frac{1}{3} , \;\; \nu =1+2n ,\;\; n \ge 0 \nonumber
\end{align}
and
\begin{align}
r^{(-n)}&=\frac{W_{n+1} \lbrack 
{\widehat F}_0^{(0)}, {\widehat F}_1^{(0)}, {\ldots} , {\widehat F}_n^{(0)}\rbrack}
{W_{n} \lbrack {\widehat F}_0^{(0)}, {\widehat F}_1^{(0)}, {\ldots} , {\widehat F}_{n-1}^{(0)}\rbrack}
\exp \left( \frac{2x^2}{3}\right),
\quad q^{(-n)}=-1/r^{(-n+1)} \lab{rponethird}\\
\mu&=\pm \frac{1}{3} , \;\; \nu =1-2n ,\;\; n \ge 0 \, ,\nonumber
\end{align}
\end{subequations}
The Wronskian identities :
\[
\begin{split}
W_{n+1} \lbrack 
F_0^{(0)}, F_1^{(0)}, {\ldots} , F_n^{(0)}\rbrack
&=-W_{n} \lbrack 
F_0^{(2)}, F_1^{(2)}, {\ldots} , F_{n-1}^{(2)}\rbrack,\\
W_{n+1} \lbrack 
{\widehat F}_0^{(0)}, {\widehat F}_1^{(0)}, {\ldots} , {\widehat F}_n^{(0)}\rbrack
&=W_{n} \lbrack 
{\widehat F}_0^{(2)}, {\widehat F}_1^{(2)}, {\ldots} , {\widehat F}_{n-1}^{(2)}\rbrack
\end{split}
\]
ensure that the solutions $y,w$ of the Painlev\'e IV equations
generated by expressions in eqs. \rf{ponethird} coincide
with those solutions $y,w$  which originate from equations \rf{monethird} 
for equal parameter $\nu$ (accomplished by shifting $n$ 
to $n\pm 1$ when going from \rf{monethird} to \rf{ponethird}).

Moreover one obtains the following Wronskian expressions
for the $\rho$-function:
\[ 
\begin{split}
\rho^{(-n)}&=G^{-n} (\rho)=
\rho + 2n \frac{4x}{3}+  
2 \left(\ln W_{n} \lbrack {\widehat F}_0^{(0)}, {\widehat F}_1^{(0)}, {\ldots} , {\widehat F}_{n-1}^{(0)}
\rbrack\right)_x, \quad n=1,2,3,{\ldots} \\
\rho^{(n)}&=G^{n} (\rho)=
\rho - 2n \frac{4x}{3}+  
2 \left(\ln W_{n} \lbrack F_0^{(2)}, F_1^{(2)}, {\ldots} , F_{n-1}^{(2)}
\rbrack\right)_x, \quad n=1,2,3,{\ldots} 
\end{split}
\] 
which satisfies the $\rho$-equation \rf{rho-eq} with $\mu^2 = \frac49$,
$\nu = 1\pm 2 n$.

Wronskians $W_{n} \lbrack F_0^{(1)} , F_1^{(1)}, {\ldots} , 
F_{n-1}^{(1)}\rbrack$ and 
$W_{n} \lbrack {\widehat F}_0^{(0)}, {\widehat F}_1^{(0)}, {\ldots} , {\widehat F}_{n-1}^{(0)}
\rbrack $ are proportional to the Okamoto polynomials
$R_n, Q_n$ as defined in \cite{Clarkson-jmp}.

\subsubsection{Solutions with $\mathbf{\mu =\pm 4/3 ,\pm 5/3 ,{\ldots}} $}

To reach expressions for solutions with $\mu=\pm 4/3, \pm 5/3,{\ldots} $
we have employed transformations $g,g^{-1}$ with their properties of raising
and lowering $\mu $ and $\nu$ by one when applied to $\sj, \bsj$ configuration.
This property of the $g,g^{-1}$ transformations ensures that we  
reach all allowed values of the Painlev\'e IV parameters 
following the zig-zag DB orbits through the $(\mu,\nu)$-plane.

We now present results obtained by applying $g,g^{-1}$ transformations
to the three basic cases with the values $\mu^2 = (2/3)^2, 
(1/3)^2$ presented in equations \rf{twothird}, \rf{monethird} 
and \rf{ponethird} in the  previous subsection.

We start with a chain of solutions \rf{twothird} to the Painlev\'e IV 
equations with $\mu^2 = (2/3)^2$. Through the successive actions 
of $g, g^{-1}$ transformations these solutions generalize to :
\begin{subequations}\label{eq:53m}
\begin{align}
y&=-\left(\ln \left(
\frac{W_{k+m+2} \bigl\lbrack 
F_0^{(1)}, F_1^{(1)}, {\ldots} , F_k^{(1)},
F_0^{(2)}, F_1^{(2)}, {\ldots} , F_m^{(2)}
\bigr\rbrack}
{W_{k+m+1} \bigl\lbrack F_0^{(1)}, F_1^{(1)}, {\ldots} , F_{k-1}^{(1)},
F_0^{(2)}, F_1^{(2)}, {\ldots} , F_{m}^{(2)}
\bigr\rbrack}\right)\right)_x-\frac{2x}{3}, \lab{53my}\\
\mu^2&=\left(\frac{2}{3}+m+1\right)^2 ,\qquad \nu=2k-m-1  \nonumber\\
w&=\left(\ln \left(
\frac{W_{k+m+2} \bigl\lbrack 
{\widehat F}_0^{(1)}, {\widehat F}_1^{(1)}, {\ldots} , {\widehat F}_k^{(1)},
{\widehat F}_0^{(2)}, {\widehat F}_1^{(2)}, {\ldots} , {\widehat F}_m^{(2)}
\bigr\rbrack}
{W_{k+m+1} \bigl\lbrack {\widehat F}_0^{(1)}, {\widehat F}_1^{(1)}, {\ldots}
, {\widehat F}_{k-1}^{(1)},
{\widehat F}_0^{(2)}, {\widehat F}_1^{(2)}, {\ldots} , {\widehat F}_{m}^{(2)}
\bigr\rbrack}\right) \right)_x-\frac{2x}{3},\lab{53mw}\\
\mu^2&=\left(\frac{2}{3}+m+1\right)^2 ,\qquad \nu =1+m-2k 
\nonumber 
\end{align}
\end{subequations}
and solve the Painlev\'e IV equations \rf{painy} and \rf{painw},
respectively, for  the positive
integers $k, m \ge 0$.

Generalizing through  the DB approach the system of solutions
given in  eqs. \rf{monethird} leads to :
\begin{subequations} \lab{43m}
\begin{align}
y&=-\left(\ln 
\left(\frac{W_{k+m+2} \bigl\lbrack 
F_0^{(1)}, F_1^{(1)}, {\ldots} , F_m^{(1)},
F_0^{(0)}, F_1^{(0)}, {\ldots} , F_k^{(0)}
\bigr\rbrack}
{W_{k+m+1} \bigl\lbrack F_0^{(1)}, F_1^{(1)}, {\ldots} , F_{m}^{(1)},
F_0^{(0)}, F_1^{(0)}, {\ldots} , F_{k-1}^{(0)}
\bigr\rbrack}\right)\right)_x-\frac{2x}{3}, \lab{43my}\\
\mu^2 &=\left(-\frac13 +m+1\right)^2=\left(\frac23 +m\right)^2, \qquad \nu=2k-m-2  \nonumber\\
w&=\left(\ln \left(
\frac{W_{k+m+2} \bigl\lbrack 
{\widehat F}_0^{(1)}, {\widehat F}_1^{(1)}, {\ldots} , {\widehat F}_m^{(1)},
{\widehat F}_0^{(0)}, {\widehat F}_1^{(0)}, {\ldots} , {\widehat F}_k^{(0)}
\bigr\rbrack}
{W_{k+m+1} \bigl\lbrack {\widehat F}_0^{(1)}, {\widehat F}_1^{(1)}, {\ldots} , {\widehat F}_{m}^{(1)},
{\widehat F}_0^{(0)}, {\widehat F}_1^{(0)}, {\ldots} , {\widehat F}_{k-1}^{(0)}
\bigr\rbrack}
\right)\right)_x-\frac{2x}{3}, \lab{43mw} \\
\mu^2 &=\left(-\frac13 +m+1\right)^2=\left(\frac23 +m\right)^2, \qquad \nu=-2k+m+2  \nonumber
\end{align}
\end{subequations}
written in terms of two positive
integers $k$ and $m$.

By applying the same procedure to solutions given
in equations \rf{ponethird} one obtains a new class of
solutions :
\begin{subequations} \lab{43pa}
\begin{align}
y&=-\left(\ln 
\left(\frac{W_{k+m+2} \bigl\lbrack 
F_0^{(1)}, F_1^{(1)}, {\ldots} , F_k^{(1)},
F_0^{(0)}, F_1^{(0)}, {\ldots} , F_m^{(0)}
\bigr\rbrack}
{W_{k+m+1} \bigl\lbrack F_0^{(1)}, F_1^{(1)}, {\ldots} , F_{k-1}^{(1)},
F_0^{(0)}, F_1^{(0)}, {\ldots} , F_{m}^{(0)}
\bigr\rbrack}\right)\right)_x-\frac{2x}{3}, \lab{43pay}\\
\mu^2 &=\left(\frac13 +m\right)^2, \qquad \nu=2k-m-1  \nonumber\\
w&=\left(\ln \left(
\frac{W_{k+m+2} \bigl\lbrack 
{\widehat F}_0^{(1)}, {\widehat F}_1^{(1)}, {\ldots} , {\widehat F}_k^{(1)},
{\widehat F}_0^{(0)}, {\widehat F}_1^{(0)}, {\ldots} , {\widehat F}_m^{(0)}
\bigr\rbrack}
{W_{k+m+1} \bigl\lbrack {\widehat F}_0^{(1)}, {\widehat F}_1^{(1)}, {\ldots}
, {\widehat F}_{k-1}^{(1)}, {\widehat F}_0^{(0)}, {\widehat F}_1^{(0)}, 
{\ldots} , {\widehat F}_{m}^{(0)}
\bigr\rbrack}
\right)\right)_x-\frac{2x}{3}, \lab{43paw} \\
\mu^2 &=\left(\frac13 +m\right)^2, \qquad \nu=m-2k+1  \nonumber
\end{align}
\end{subequations}
written in terms of two positive
integers $k,m \ge 0$.

It is important to point out that the first two of the above three 
classes of solutions overlap in their values of the $\mu$ and $\nu$ 
parameters for $\mu^2 > (2/3)^2$ and in such cases the corresponding
solutions coincide. For instance, solutions \rf{53my} with
$\mu^2=(2/3+m+1)^2 ,\,\nu=2k-m-1$ are equal to solutions \rf{43my} with
$\mu^2 =(-1/3 +m^{\prime}+1)^2,\, \nu=2k^{\prime}-m^{\prime}-2$
when $m^{\prime}=m+1$ and $k^{\prime}= k+1$. 
Thus, for any solution $y_{m, k}$ or $w_{m,k}$ from \rf{53m} there exists an identical 
solution $y_{m^{\prime}=m+1, k^{\prime}=k+1}$ or $w_{m^{\prime}=m+1, k^{\prime}=k+1}$
from \rf{43m}. As encountered before in the text, an equality between 
two classes of Painlev\'e IV solutions derived by action of 
the DB transformations stems from existence 
of a special Wronskian identity. In this case the relevant 
identity is given by :
\[
\begin{split}
&W_{k+m+2} \bigl\lbrack 
F_0^{(1)}, F_1^{(1)}, {\ldots} , F_k^{(1)},
F_0^{(0)}, F_1^{(0)}, {\ldots} , F_m^{(0)}
\bigr\rbrack= \\
&=K_{k,m} W_{k+m+4} \bigl\lbrack 
F_0^{(1)}, F_1^{(1)}, {\ldots} , F_{m+1}^{(1)},
F_0^{(0)}, F_1^{(0)}, {\ldots} , F_{k+1}^{(0)}
\bigr\rbrack\, ,
\end{split}
\]
with certain combinatorial constants $K_{k,m}$.

\section{Conclusions and Outlook}
We presented here a systematic and self-contained derivation of 
rational solutions to the Painlev\'e IV equation using the method 
of the Darboux-B\"acklund transformations of the particular
reduction of the AKNS pseudo-differential Lax hierarchy.
By studying the orbits of the Darboux-B\"acklund transformations 
originating from few seeds solutions we were able to find closed
expressions for solutions associated to all allowed values of 
the Painlev\'e IV parameters.
The explicit expressions for all Wronskian representations derived 
here seem to be in general agreement with 
determinant formulas obtained earlier in \cite{kajiwara,noumi,Clarkson}
by alternate methods.

In a separate publication we will discuss the Darboux-B\"acklund
transformations versus the (affine) Weyl group symmetry of 
the Painlev\'e IV equation \cite{okamoto}.

\vskip .4cm \noindent
{\bf Acknowledgements} \\
JFG and AHZ thank CNPq and FAPESP  for 
financial support.
Work of HA is partially supported by grant NSF PHY-0651694.

\end{document}